\documentclass[conference,letterpaper]{IEEEtran}

\usepackage[letterpaper,
  left=0.63in,right=0.63in,
  top=0.75in,bottom=1.05in
]{geometry}

\usepackage{graphicx}
\usepackage{booktabs}
\usepackage{orcidlink}

\newcommand{\System}{\textsc{NanoEdgeGuard}}

\begin{document}
\title{Adaptive Security at the Edge for 6G-Enabled Healthcare IoT}

\author{\IEEEauthorblockN{Ijaz Ahmad\orcidlink{0000-0002-6152-8947}}
\IEEEauthorblockA{Centre for Wireless Communications\\
University of Oulu, Finland\\
Email: ahmad.ijaz@oulu.fi}
\and
\IEEEauthorblockN{Ijaz Ahmad\orcidlink{0000-0003-1101-8698}}
\IEEEauthorblockA{VTT Technical Research\\
Centre of Finland, Finland\\
Email: ijaz.ahmad@vtt.fi}
\and
\IEEEauthorblockN{Erkki Harjula\orcidlink{0000-0001-5331-209X}}
\IEEEauthorblockA{Centre for Wireless Communications\\
University of Oulu, Finland\\
Email: erkki.harjula@oulu.fi}}
\maketitle

\begin{abstract}
Healthcare IoT services increasingly rely on edge gateways to relay routine telemetry and deliver rare but time-critical alarms. Even short traffic bursts can inflate worst-case delay and interfere with urgent messages. We present \System, a kernel-plane closed-loop controller that observes per-source traffic intensity at the edge and enforces an auditable, multi-tier rate policy using in-kernel traffic-control hooks. Unlike static firewall rules or user-space control loops, our design prioritizes fast actuation and explicit recovery through hysteresis, and it records policy transitions for auditability. Using a Raspberry Pi gateway hosting an MQTT broker and two ESP32 endpoints generating vitals, alarms, and a timed burst, we show that adaptive kernel-plane rate control reduces the 99th-percentile alarm RTT by \emph{13.3\%} compared to a user-space firewall baseline while maintaining no-enforcement-level RTT, and it reduces excess admitted burst traffic by \emph{46\%} compared to no enforcement. These early results indicate that adaptive, auditable enforcement at the gateway can improve resilience for healthcare IoT, and it can be extended toward on-demand policy deployment in future edge intelligence.
\end{abstract}


\IEEEpeerreviewmaketitle
\vspace{-15pt}
\section{Introduction}
\vspace{-5pt}
Healthcare is becoming more connected and more data-driven, with sensing and decision support extending beyond hospital walls into homes, ambulances, and community care. Wearables, bedside monitors, and home devices can stream vital signs continuously, while also generating rare alarms that require timely delivery. A common example is remote monitoring of high-risk patients, where routine telemetry can tolerate small fluctuations but an alarm often cannot. If alarms arrive late, triage slows down and confidence in the overall system decreases. This motivates 6G visions that place health services on an edge--cloud continuum, where computation and security functions can be positioned close to patients while still integrating with higher-tier services \cite{ahad2023sixghealth}.

In practice, the first-hop gateway plays a central role in this continuum. It often runs on a lightweight and heterogeneous edge node, ranging from a single-board computer to a feature-rich smartphone and, in some deployments, an on-premises edge server or MEC node in a private 5G environment. The gateway may host an MQTT broker and local preprocessing, and it must operate under tight CPU and memory budgets \cite{ijaz2024lightweight}. Under these constraints, worst-case delay often matters more than the average. Short congestion spikes and temporary CPU stalls can postpone urgent messages. Bursty traffic from a rogue device can further amplify contention at the gateway.

Kernel programmability offers an alternative path for fast and predictable enforcement. Prior work shows that eBPF can accelerate firewall-like processing and improve dataplane efficiency \cite{bertrone2018bpfiptables}. Other studies emphasize that in-kernel offload is most beneficial when the task remains small and carefully designed \cite{hohlfeld2019xdp,shahinfar2025demystifying}. These results motivate a healthcare-centric focus that goes beyond throughput. The core need is stable protection of latency-sensitive messages, together with a clear record of decisions that supports post-event analysis.

\begin{figure}[htbp]
  \centering
  \includegraphics[width=\linewidth]{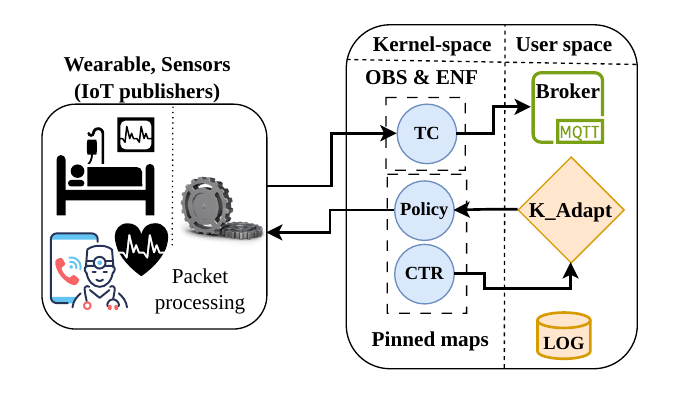}
  \vspace{-30pt}
  \caption{\System ~on a nano-edge gateway. The controller adapts rate-limit dynamic policy and enforces them in the kernel.}
  \vspace{-10pt}
  \label{fig:arch}
\end{figure}

\noindent\textbf{Contribution.} This extended abstract presents \System, an adaptive controller for resource-constrained edge gateways that protects alarm delivery under intermittent burst overlap. It enforces tiered rate control in kernel space using TC/eBPF and records policy transitions in structured logs. Fig.~\ref{fig:arch} summarizes the architecture, and Fig.~\ref{fig:results} reports the initial tail-delay and containment results on a Raspberry Pi gateway with MQTT and ESP32 endpoints.
\vspace{-20pt}
\vspace{-10pt}
\section{System Design and Prototyping}
\System ~follows an observe-decide-enforce loop at the gateway (Fig.~\ref{fig:arch}). The observe stage collects per-device traffic intensity from kernel counters and a lightweight service-quality signal from the broker. The decision stage selects a rate-limit tier using a simple risk score and hysteresis. Hysteresis includes persistence and a hold-down period so that escalation and recovery are explicit and stable under intermittent stress. The enforce stage applies the selected tier close to the packet path using TC/eBPF and pinned policy maps. The pinned state keeps enforcement stable across process restarts while allowing user-space logic to remain auditable.

The controller selects among four actions: \textsc{Monitor}, \textsc{Rate-Low}, \textsc{Rate-Mid}, and \textsc{Rate-High}. Hard drop is reserved for extreme or non-session-critical flows. For TCP-based MQTT telemetry, sustained drops can trigger timeouts and reconnect loops, which confounds fair quality-of-service evaluation. We therefore use tiered rate control as the default policy in this work. Auditability is a design goal. Each control interval emits a structured JSONL record including observed signals, selected tier, applied tier, and a transition reason. These logs make policy evolution and recovery verifiable across repeated runs.
\vspace{-10pt}
\section{Experimental Setup}
\textbf{Testbed.} A Raspberry Pi gateway runs a Mosquitto broker and the \System ~controller. Two ESP32 endpoints publish vitals and alarms as listed in Table \ref{tab:workload}. 

\textbf{Stress model.} We model a misbehaving (or misconfigured) endpoint as an MQTT publisher that increases its publish intensity in short ON/OFF pulses during a fixed window (UE-B, 60--120\,s). This creates bursty contention at the gateway and broker while using the same messaging stack as legitimate traffic. The goal is to contain these pulses quickly while preserving timely alarm delivery from UE-A.

\textbf{Cases.} We compare \textsc{No\_Enf} (no enforcement), \textsc{U\_NFT} (user-space firewall baseline), and \textsc{K\_Adapt} (kernel-plane adaptive tiers). Each case is repeated three times.

\textbf{Metrics.} We measure alarm acknowledgment delay using MQTT QoS1 PUBACK round-trip time and report p95/p99 as median-of-3. We measure burst containment as flood-window excess admitted bytes (60--120\,s) computed relative to the benign baseline estimated from 0--60\,s, also as median-of-3. Adaptivity is summarized via tier transitions from logs.

\begin{table}[htbp]
\centering
\caption{Endpoints and stress workload.}
\label{tab:workload}
\vspace{-5pt}
\begin{tabular}{@{}lll@{}}
\toprule
\textbf{Node} & \textbf{Benign traffic} & \textbf{Burst behavior} \\
\midrule
UE-A & vitals + alarm bursts & -- \\
UE-B & slow-rate vitals & ON/OFF publish pulses (60--120\,s) \\
Gateway & Mosquitto + controller & enforcement: No\_Enf / U\_NFT / K\_Adapt \\
\bottomrule
\end{tabular}
\vspace{-10pt}
\end{table}

\section{Results}
\begin{figure}[htpb]
  \centering
  \includegraphics[width=\linewidth]{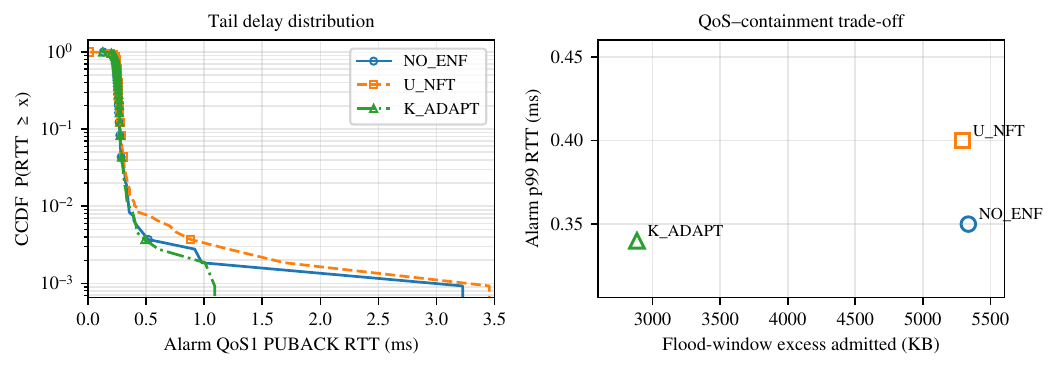}
  \vspace{-20pt}
  \caption{(a) Tail CCDF of alarm QoS1 PUBACK RTT under pulse-flood stress (log-y; x$\leq$3.5 ms). (b) QoS–containment trade-off: alarm p99 RTT vs flood-window excess admitted bytes (60–120 s, excess over 0–60 s baseline).}
  \vspace{-15pt}
  \label{fig:results}
\end{figure}

Fig.~\ref{fig:results}(a) shows the tail CCDF of alarm acknowledgment delay under pulse-burst stress. Compared to the user-space baseline, \textsc{K\_Adapt} reduces the alarm p99 delay from 0.398\,ms to 0.345\,ms, which is a 13.3\% improvement (median-of-3). The \textsc{K\_Adapt} p99 remains close to the no-enforcement baseline (0.348\,ms), indicating that adaptive rate control protects tail behavior avoiding overhead that dominates latency.

Fig.~\ref{fig:results}(b) summarizes the quality--containment trade-off using alarm p99 delay and flood-window excess admitted bytes. Under \textsc{No\_Enf}, excess admitted bytes are 5336.5\,KB. \textsc{K\_Adapt} reduces this to 2884.1\,KB, a 46.0\% reduction (median-of-3). The controller escalates during ON pulses and relaxes during OFF periods, which is consistent with the hysteresis-based recovery design and avoids rapid oscillation.

The U\_NFT baseline applies firewall-style controls through periodic user-space updates. Under bursty ON/OFF behavior, coarse update timing can react after a pulse has already increased contention, and frequent rule updates can add control overhead when the gateway is stressed. In contrast, \textsc{K\_Adapt} maintains per-source state in pinned maps and adjusts tiers with explicit hysteresis, which supports faster and smoother containment during pulses and cleaner recovery afterward. This scenario is challenging for rule-based firewalls because benign and misbehaving publishers use the same broker service, so simple static rules have limited context about messaging-level overload. Our controller uses a lightweight service-quality signal and hysteresis-driven tiers, which helps it react to pulses and recover smoothly.

\section{Conclusion and Future Work}
We presented \System, an adaptive and auditable kernel-plane rate controller for healthcare IoT gateways. In a pulse-burst workload, \textsc{K\_Adapt} improves alarm p99 delay by 13.3\% over a user-space firewall baseline while reducing excess admitted burst traffic by 46.0\% over no enforcement. These early results support kernel-plane adaptive rate control as a practical mechanism for latency-sensitive healthcare IoT on constrained edge devices. Current results focus on two endpoints and a single gateway under a pulse-burst workload. Broader multi-source contention, stealthier abuse patterns, and cross-tier escalation beyond the gateway are left for future runs. A next step is to integrate on-demand policy deployment, where lightweight enforcement components are activated when and where risk rises in the distributed edge-cloud continuum.

\vspace{-5pt}
\section*{Acknowledgment}
\noindent This research is supported by the Finnish Doctoral Program Network in Artificial Intelligence, AI-DOC (decision number VN/3137/2024-OKM-6) and Research Council of Finland funded projects 6G Flagship (346208) and Profi6 (336449).

\bibliographystyle{IEEEtran}
\bibliography{references}

\end{document}